\documentclass[twocolumn,twoside,aps,prl,showpacs,superscriptaddress]{revtex4-1}
\usepackage{graphicx}
\usepackage{amsmath,amssymb}
\usepackage{color,revsymb4-1}

\newcommand{\1}{{\openone}}
\newcommand{\qed}{{\hfill$\Box$}}
\newcommand{\tr}{{\operatorname{tr}\,}}

\begin{document}

\title{(Non-)Contextuality of Physical Theories as an Axiom}

%%%%%%%%%%%%%%%%%%%%%%%%%%%%%%%%%%%%%%%%%%%%%%%%%%%%%%%%%%%%%%%%%%%

\author{Ad\'an Cabello}
 \email{adan@us.es}
 \affiliation{Departamento de F\'{\i}sica Aplicada II, Universidad de Sevilla, E-41012 Sevilla, Spain}

\author{Simone Severini}
 \email{simoseve@gmail.com}
 \affiliation{Department of Physics \& Astronomy, University College London, WC1E 6BT London, United Kingdom}

 \author{Andreas Winter}
 \email{a.j.winter@bris.ac.uk}
 \affiliation{Department of Mathematics, University of Bristol, Bristol BS8 1TW, United Kingdom}
 \affiliation{Centre for Quantum Technologies, National University of Singapore, 2 Science Drive 3, Singapore 117542}

%%%%%%%%%%%%%%%%%%%%%%%%%%%%%%%%%%%%%%%%%%%%%%%%%%%%%%%%%%%%%%%%%%%

\date{11 October 2010}

%First version: August 17, 2010 (AC).
%Second version: August 23, 2010 (SS).
%Third version: 16 September 2010 (AW).
%Fourth version: 8 October 2010 (AW)
%Fifth version: 9 October 2010 (AC)
%This: 11 October 2010 (Aw)

%%%%%%%%%%%%%%%%%%%%%%%%%%%%%%%%%%%%%%%%%%%%%%%%%%%%%%%%%%%%%%%%%%%

\begin{abstract}
We show that the noncontextual inequality proposed by Klyachko
{\em et al.} [Phys. Rev. Lett. {\bf 101}, 020403 (2008)]
belongs to a broader family of inequalities, one associated to
each compatibility structure of a set of events (a graph), and
its independence number. These have the surprising property
that the maximum quantum violation is given by the Lov\'{a}sz
$\vartheta$-function of the graph, which was originally
proposed as an upper bound on its Shannon capacity.
Furthermore, probabilistic theories beyond quantum mechanics
may have an even larger violation, which is given by the
so-called fractional packing number. We discuss in detail, and
compare, the sets of probability distributions attainable by
noncontextual, quantum, and generalized models; the latter two
are shown to have semidefinite and linear characterizations,
respectively. The implications for Bell inequalities, which are
examples of noncontextual inequalities, are discussed. In particular,
we show that every Bell inequality can be recast as a noncontextual
inequality \`{a} la Klyachko \emph{et al.}
\end{abstract}
%%%%%%%%%%%%%%%%%%%%%%%%%%%%%%%%%%%%%%%%%%%%%%%%%%%%%%%%%%%%%%%%%%%

%Entanglement and quantum nonlocality
%(e.g. EPR paradox, Bell's inequalities, GHZ states, etc.)
%Quantum information
%graph theory
\maketitle

%%%%%%%%%%%%%%%%%%%%%%%%%%%%%%%%%%%%%%%%%%%%%%%%%%%%%%%%%%%%%%%%%%%

%\medskip%\noindent
{\it Introduction.}---Recently, Klyachko, Can, Binicio\u{g}lu,
and Shumovsky (KCBS) \cite{KCBS08} have introduced a
noncontextual inequality (\emph{i.e.}, one satisfied by any
noncontextual hidden variable theory), which is violated by
quantum mechanics, and therefore can be used to detect quantum
effects. The simplest physical system which exhibits quantum
features in this sense is a three-level quantum system or
qutrit \cite{Gleason57, Bell66, KS67}. The KCBS inequality is
the simplest noncontextual inequality violated by a qutrit, in
a similar way that the Clauser-Horne-Shimony-Holt (CHSH)
inequality \cite{CHSH69} is the simplest Bell inequality
violated by a two-qubit system.

The KCBS inequality has been recently tested in the laboratory
\cite{Lapkiewicz10} and has stimulated many recent developments
\cite{Cabello08, BBCP09, KZGKGCBR09, ARBC09, Cabello10,
GKCLKZGR10, Cabello10b}. It can adopt two equivalent forms.
Consider $5$ \emph{yes-no} questions $P_{i}$ ($i=0,\ldots,4$)
such that $P_{j}$ and $P_{j+1}$ (with the sum modulo $5$) are
\emph{compatible}: both questions can be jointly asked without
mutual disturbance, so, when the questions are repeated, the
same answers are obtained; and \emph{exclusive}: not both can
be true. One can represent each of these questions as a vertex
of a pentagon (\emph{i.e.}, a $5$-cycle) where the edges denote
compatibility and exclusiveness. What is the maximum number of
\emph{yes} answers one can get when asking the $5$ questions to
a physical system? Clearly, two, because of the exclusiveness
condition \cite{Wright78}. If we denote \emph{yes} and
\emph{no} by $1$ and $0$, respectively, then, even if one asks
only one question to each one of an identically prepared
collection of systems, and then count the average number of
\emph{yes} answers corresponding to each question, the
following inequality holds:
\begin{equation}
  \beta := \sum_{i=0}^{4}\langle P_{i}\rangle\leq2, \label{KCBS1}
\end{equation}
if we assume that these answers are predetermined by a hidden variable theory.
This is the first form of
the KCBS inequality. What has \eqref{KCBS1} to do with noncontextuality?
Noncontextual hidden variable theories are those in which the answer of $P_{j}$ is
independent of whether one ask $P_{j}$ together with $P_{j-1}$ (which is
compatible with $P_{j}$), or together with $P_{j+1}$ (which is also compatible
with $P_{j}$). A set of mutually compatible questions is called a
\emph{context}. Since, $P_{j+1}$ and $P_{j-1}$ are not necessarily
compatible, $\{P_{j},P_{j-1}\}$ is one context and $\{P_{j},P_{j+1}\}$ is a
different one, and they are not both contained in a joint context.
The assumption is that the answer to $P_{j}$ will be the same in both.

Now, let us consider contexts instead of questions, \emph{i.e.}, let us ask
individual systems not one but two compatible and exclusive questions. In the
pentagon, a context is represented by an edge connecting two vertices, so we
have $5$ different contexts. In order to study the correlations between the
answers to these questions, it is useful to transform each question into a
dichotomic observable with possible values $-1$ (no) or $+1$ (yes), so when both
questions give the same answer the product of the results of the observables
is $+1$, but when the answers are different then the product of the results of
the observables is $-1$. For instance, this can be done by defining the
observables $A_{i}=2P_{i}-1$. Then, inequality \eqref{KCBS1} is equivalent to
the noncontextual correlation inequality, the second form of KCBS,
\begin{equation}
  \beta^{\prime} := \sum_{i=0}^{4}\langle A_{i}A_{i+1}\rangle \geq -3,
  \label{KCBS}
\end{equation}
which can be derived independently based solely on the
assumption that the observables $A_{i}$ have noncontextual
results $-1$ or $+1$. \emph{I.e.}, we do not need to assume
exclusiveness to derive it, effectively because the occurrence
of correlation functions $\langle A_i A_{i+1} \rangle$
implements a penalty for violating exclusiveness.

For a qutrit, the maximum quantum violation of inequality
\eqref{KCBS1} was shown to be $\beta_{\mathrm{QM}}(5) =
\sqrt{5} \approx 2.236$, which is equivalent to a violation of
inequality \eqref{KCBS} of $\beta_{\mathrm{QM}}^{\prime}(5) =
5-4\sqrt{5} \approx -3.94.$ The maximum violation of the KCBS
inequality occurs for the state $\langle\psi|=(0,0,1)$ and the
questions $P_{i} = |v_{i}\rangle\!\langle v_{i}|$ or the
observables $A_{i} = 2|v_{i}\rangle\!\langle v_{i}|-\openone$,
where
\begin{align}
  \langle v_{0}|   &= N_{0}\left(1,0,\sqrt{\cos{(\pi/5)}}\right), \nonumber \\
  \langle v_{1,4}| &= N_{1}\left(\cos{(4\pi/5)},\pm\sin{(4\pi/5)},\sqrt{\cos{(\pi/5)}}\right),\\
  \langle v_{2,3}| &= N_{2}\left(\cos{(2\pi/5)},\mp\sin{(2\pi/5)},\sqrt{\cos{(\pi/5)}}\right), \nonumber
\end{align}
the $N_{i}$ being suitable normalization factors. These vectors
connect the origin with the vertices of a regular pentagon.
Interestingly, with this choice, $\langle
A_{i}A_{i+1}\rangle=\left[-1+3\cos\left(\pi/5\right) \right]
{\sec ^{2}\left(\pi/10\right)/2}$, for $i=0,\ldots,n-1$.
Observe that $\langle v_{i}|v_{i+1}\rangle=0$ and
$\beta_{\mathrm{QM}}(5)=\sum_{i\operatorname{mod}
5}|\langle\psi|v_{i}\rangle|^{2}$. The vectors that give
$\beta_{\mathrm{QM}}(5)$ form an orthonormal representation of
the $5$-cycle.

\medskip%\noindent
{\it General compatibility structures.}---
The KCBS inequality suggests itself a generalization to arbitrary
graphs instead of the pentagon.
Most generally and abstractly, Kochen-Specker (KS) theorems
\cite{KS67} are about the possibility of interpreting a given
structure of compatibility of ``events,'' and additional
constraints such as exclusiveness, in a classical or
nonclassical probabilistic theory. In this paper, these events
are interpreted as \emph{atomic} events, each of which can
occur in different contexts. Formally, the events are labelled
by a set $V$ (in practice finite, and often just integer
indices, $V=\{1,2,\ldots,n\}$). The set of all valid contexts
is a \emph{hypergraph} $\Gamma$, which is simply a collection
of subsets $C \subset V$; note that for hypergraphs of
contexts, with each $C\in \Gamma$, all of the subsets of $C$
are also valid contexts, and hence part of $C$.
The interpretation is that there should exist (deterministic) events in a
probabilistic model, one $P_i$ for each $i\in V$, and for each context $C$ a
measurement among whose outcomes are the $P_i$ ($i\in C$). The events are
hence mutually exclusive, as in the measurement postulated to exist
for some $C\in\Gamma$, at most one outcome $i\in C$ can occur.
For instance, a \emph{classical (noncontextual) model} would be a measurable
space $\Omega$, with each $P_i$ being the indicator function of
a measurable set (an event, in fact) such that for all $C\in\Gamma$,
$\sum_{i\in C} P_i \leq 1$ (\emph{i.e.}, the supporting sets of the
$P_i$ should be pairwise disjoint).

In contrast, a \emph{quantum model} requires a Hilbert space ${\cal H}$
and associates projection operators $P_i$ to all $i\in V$,
such that for all $C\in\Gamma$, $\sum_{i\in C} P_i \leq \1$
(\emph{i.e.}, the $P_i$ can be thought of as outcomes in a von
Neumann measurement).

Thanks to KS we know that quantum models are strictly more
powerful that classical ones; but they are still not the most
general ones. A \emph{generalized model} requires choosing a
generalized probabilistic theory in which the $P_i$ can be
interpreted as measurement outcomes: following
\cite{Mackey,Ludwig-1,Ludwig-2,Holevo-book,Barrett07}, formally
it consists of a real vector space ${\cal A}$ of observables,
with a distinguished unit element $u\in{\cal A}$ and a vector
space order: the latter is given by the closed convex cone
${\cal P} \subset {\cal A}$ of positive elements containing $u$
in its interior, such that ${\cal P}$ spans ${\cal A}$ and is
pointed, meaning that, with the exception of $0$, ${\cal P}$ is
entirely on one side of a hyperplane. For two elements $X,Y \in
{\cal A}$ we then say $X \leq Y$ if and only if $Y-X \in {\cal
P}$. (We shall only discuss finite dimensional ${\cal A}$,
otherwise there will be additional topological requirements.)
The elements with $0 \leq E \leq u$ are called \emph{effects}.
This structure is enough to talk about measurements: they are
collections of effects $(E_1,\ldots,E_k)$ such that
$\sum_{j=1}^k E_j = u$.

[Observe how we recover quantum mechanics when ${\cal P}$ consists
of the semidefinite matrices within the Hermitian ones over a
Hilbert space, and $u=\1$. Classical probability instead, when
${\cal P}$ are the non-negative functions within the measurable
ones over a measure space, $u$ being the constant $1$ function.]

Now, a generalized model for the hypergraph $\Gamma$ is the
association of an effect $P_i \in {\cal A}$ to each $i\in V$, such that
each $P_i$ is a sum of normalized extremal effects, and
for all $C\in\Gamma$, $\sum_{i\in C} P_i \leq u$. The latter condition
ensures that the family $(P_i:i\in C)$ can be completed to a
measurement, possibly in a larger space $\widetilde{\cal A} \supset {\cal A}$.
We finally demand that this can be done such that also
$u-\sum_{i\in C} P_i$ is a sum of normalized extremal effects.

Notice that in all of the above we never require that any
particular context should be associated to a complete
measurement: the conditions only make sure that each context is
a subset of outcomes of a measurement and that they are
mutually exclusive. Thus, unlike the original KS theorem, it is
clear that every context hypergraph $\Gamma$ has always a
classical noncontextual model, besides possibly quantum and
generalized models. This is where noncontextual inequalities
come in: note that all of the above types of models allow for
the choice of a state (be it a probability density, a quantum
density operator, or generalized state), under which all
expectation values $\langle P_i \rangle$ make sense, and hence
also the expression
\begin{equation}
  \beta = \sum_{i\in V} \langle P_i \rangle.
\end{equation}
Moreover, all probabilities $\langle P_i \rangle$ are independent of
the context in which $P_i$ occurs, as they depend only on the
effect $P_i$ and the underlying state.
Since this is the condition underlying Gleason's theorem, we call it
the \emph{Gleason property}.

We can then ask for the set of all attainable vectors
$\bigl(\langle P_i \rangle\bigr)_{i\in V}$
for given hypergraph $\Gamma$, over all models of a given sort
(classical noncontextual, quantum mechanical, or generalized
probabilistic theory) and states within it. These are evidently
convex subsets in $[0,1]^V \subset \mathbb{R}^V$; we denote
the sets of noncontextual, quantum and generalized expectations by
$\mathcal{E}_\mathrm{C}(\Gamma)$, $\mathcal{E}_\mathrm{QM}(\Gamma)$
and $\mathcal{E}_\mathrm{GPT}(\Gamma)$, respectively.
The central task of the present theory is to characterize
these convex sets and to compare them for various $\Gamma$.
This is because a point $\vec{p} \in \mathcal{E}_\mathrm{X}(\Gamma)$
in any of these sets describes the outcome probabilities of any
compatible set of events (\emph{i.e.}, any context). Note that
all of them are \emph{corners} in the language of \cite{Knuth94}:
if $0 \leq q_i \leq p_i$ for all $i \in V$, then
$\vec{p} \in \mathcal{E}_\mathrm{X}(\Gamma)$ implies also
$\vec{q} \in \mathcal{E}_\mathrm{X}(\Gamma)$.

In particular, the extreme values of $\beta$ over these sets
are denoted $\beta_{\mathrm{C}}(\Gamma)$,
$\beta_{\mathrm{QM}}(\Gamma)$, and
$\beta_{\mathrm{GPT}}(\Gamma)$, respectively. It is clear that
\begin{equation}
  \beta_{\mathrm{C}}(\Gamma)
    \leq \beta_{\mathrm{QM}}(\Gamma)
    \leq \beta_{\mathrm{GPT}}(\Gamma)
\end{equation}
by definition.

\medskip%\noindent
{\it Maximum values.}---Prepared by the
above discussion,
for given hypergraph $\Gamma$, we can define the adjacency
graph $G$ on the vertex set $V$:
two $i,j\in V$ are joined by an edge if and only if there exists
a $C\in \Gamma$ such that both $i,j \in C$. Then,
\begin{equation}
  \label{eq:beta:C-QM}
  \beta_{\mathrm{C}}(\Gamma) = \alpha(G),
  \quad
  \beta_{\mathrm{QM}}(\Gamma) = \vartheta(G),
\end{equation}
where $\alpha(G)$ is the independence number of the graph,
\emph{i.e.} the maximum number of pairwise disconnected
vertices, and $\vartheta(G)$ is the Lov\'{a}sz
$\vartheta$-function of $G$ \cite{Lovasz79, Knuth94, Korner98},
defined as follows: First, an \emph{orthonormal representation}
(OR) of a graph is a set of unit vectors associated to the
vertices such that two vectors are orthogonal if the
corresponding vertices are adjacent. Then,
\begin{equation}
  \vartheta(G) := \max \sum_{i=1}^{n} |\langle\psi|v_{i}\rangle|^{2},
  \label{lov}
\end{equation}
where the maximum is taken over all unit vectors $|\psi\rangle$ (in Eucledian
space) and ORs $\{|v_{i}\rangle:i=1,\ldots,n\}$ of ${G}$ \cite{Lovasz09}.
Note that on the right hand side, we can get rid of $|\psi\rangle$
by observing
\begin{equation}
  \max_{|\psi\rangle} \sum_{i=1}^{n} |\langle\psi|v_{i}\rangle|^{2}
     = \left\| \sum_{i=1}^n |v_i\rangle\!\langle v_i| \right\|_\infty.
\end{equation}
Furthermore, $\vartheta(G)$ is given by a semidefinite program \cite{Lovasz79},
which explains the key importance of this number for combinatorial
optimization and zero-error information theory -- indeed $\vartheta(G)$
is an upper bound to the Shannon capacity of a graph \cite{Lovasz79}.

Observe that this says in particular that when discussing classical and
quantum models, we never need to consider contexts of more than
two events. Indeed, it is a (nontrivial) property of these models that
if in a set of events any pair is compatible and exclusive, then
so is the whole set; more generalized probabilistic theories do not
have this property, cf.~\cite{LSW:seer}.

To prove Eq.~(\ref{eq:beta:C-QM}), we notice that for a given
probabilistic model, the expectation is always maximized on an
extremal, \emph{i.e.}~pure, state. In the classical case, this
amounts to choosing a point $\omega \in \Omega$, so that $w_i
:= P_i(\omega)$ is a $0$-$1$-valuation of the set $V$. By
definition, it has the property that, in each hyperedge
$C\in\Gamma$, at most one element is marked $1$, and $\beta$ is
simply the number of marked elements. It is clear that the
marked elements form an independent set in $\Gamma$ (and
equivalently in the graph $G$). In the quantum case, let the
maximizing state be given by a unit vector $|\psi\rangle$, and
for each $i$, $\langle \psi | P_i | \psi \rangle = |\langle
\psi | v_i \rangle|^2$, for $|v_i\rangle := P_i|\psi\rangle /
{\sqrt{\psi | P_i | \psi \rangle}}$. This clearly is an
orthogonal representation of $G$, in fact the projectors
$|v_i\rangle\!\langle v_i|$ form another quantum model of
$\Gamma$, with the same maximum value of $\beta$, which by the
definition we gave earlier is just Lov\'{a}sz' $\vartheta(G)$.

Each graph $G$ where $\alpha(G) < \vartheta(G)$ thus exhibits a
limitation of classical noncontextuality, which can be
witnessed in experiments with an appropriate set of projectors,
and on an appropriate state. In this sense, each such graph
provides a proof of the KS theorem.

Taking $n\geq 5$ odd and applying a result from \cite{Lovasz79} to $G=C_n$,
the $n$-cycle, one obtains the noncontextual quantum bounds
\begin{equation}
  \beta_{\mathrm{QM}}(n) = \vartheta(C_{n})
                         = \frac{n\cos\left(\pi/n\right)}{1+\cos\left(\pi/n\right)},
\end{equation}
where $C_{n}$ denotes the $n$-cycle. After some algebra, the quantum
bound for the analogue of (\ref{KCBS}) can be written as
\begin{equation}
  \beta_{\mathrm{QM}}^{\prime}(n)
         = \frac{n}{2}\left[  -1+3\cos\left(\frac{\pi}{n}\right)\right]
                 {\sec^{2}\left(\frac{\pi}{2n}\right)},
  \label{Tsirelsonodd}
\end{equation}
for all state space dimensions larger or equal to $3$; the same
result was obtained recently by Liang, Spekkens, and
Wiseman~\cite{LSW:seer}.
%To give an extra example, when $n=7$, taking
%\begin{equation}
%\begin{tabular}[c]{l}
%$\langle v_{0}|=M_{1}\left(1,0,\sqrt{-\cos{(6\pi/7)}}\right),$\\
%$\langle v_{1,6}|=M_{1,6}\left(\cos{(6\pi/7)},\pm\sin{(6\pi/7)},\sqrt
%{-\cos{(6\pi/7)}}\right),$\\
%$\langle v_{2,5}|=M_{2,5}\left(\cos{(2\pi/7)},\mp\sin{(2\pi/7)},\sqrt
%{-\cos{(6\pi/7)}}\right),$\\
%$\langle v_{3,4}|=M_{3,4}\left(\cos{(4\pi/7)},\pm\sin{(4\pi/7)},\sqrt
%{-\cos{(6\pi/7)}}\right),$
%\end{tabular}
%\end{equation}
%where $M_{i}$ are normalization factors, one obtains
%$\beta_{\mathrm{QM}}(7) = 7\cos\left(\pi/7\right)/(1+\cos\left(\pi/7\right))$.
%
%When $\beta(G)=\vartheta(G)$ there is no quantum mechanical violation
%of classical contextuality.

We remark here that there are also ``state-independent'' KS
proofs \cite{KS67, Peres93, CEG96}: these are given by quantum
noncontextual models of a graph $G$ such that $\sum_i \langle
P_i \rangle > \alpha(G)$ for every state. The proofs in the
literature typically have this property, as they are based on
rank-one $P_i = |v_1\rangle\!\langle v_i|$, and for each $j \in
V$ there exists $C \in \Gamma$ such that $j\in C$ and
$\sum_{i\in C} P_i = \1$ (\emph{i.e.}, each $P_j$ is part of a
context that is already a complete measurement; the
$|v_i\rangle$ forming a complete orthonormal basis). Due to the
symmetric structure of most KS proofs, $\sum_i P_i$ turns out
to be proportional to the identity, so $\beta$ is independent
of the state.

It is known that $\vartheta(G)$ can be much larger than $\alpha(G)$;
in particular, it is known that (for appropriate, arbitrarily large $n$)
there are graphs $G$
with $\vartheta(G) \approx \sqrt{n}$ but $\alpha(G) \approx 2\log n$,
and others with $\vartheta(G) \approx \sqrt[4]{n}$ but $\alpha(G) = 3$
\cite{Peeters}. Hence, the quantum violation of noncontextual
inequalities can be arbitrarily large.
%e-assisted channel codes, etc...

\medskip%\noindent
{\it Description of the probability sets.}---We now show
that arbitrary linear functions can be optimized over
$\mathcal{E}_{\mathrm{QM}}(\Gamma)$ as semidefinite programs: for
an arbitrary vector $\vec{\lambda} \in R^V$, let
\begin{equation}
  \vec{\lambda}(\mathcal{E}_{\mathrm{QM}}(\Gamma))
       = \max \sum_i \lambda_i p_i
       \text{ s.t. }\vec{p} \in \mathcal{E}_{\mathrm{QM}}(\Gamma).
\end{equation}
First of all, without loss of generality, all $\lambda_i$ are non-negative;
this follows because $\mathcal{E}_{\mathrm{QM}}(\Gamma)$ is a corner
and hence $\vec{\lambda}(\mathcal{E}_{\mathrm{QM}}(\Gamma))$
is unchanged when we replace all negative
$\lambda_i$ by $0$. Now recall that $p_i = |\langle \psi | v_i \rangle|^2$
for some unit vector $|\psi\rangle$ and an orthonormal
representation $\{ |v_i\rangle \propto P_i | \psi \rangle \}$ of $G$.
Hence,
\begin{equation}\begin{split}
  \vec{\lambda}\vec{p}
     &= \sum_{i\in V} \lambda_i \langle \psi | P_i | \psi \rangle \\
     &= \langle \psi | \left( \sum_{i\in V} \lambda_i |v_i\rangle\!\langle v_i| \right) | \psi \rangle \\
     &= \langle t | \left( \sum_{ij\in V} \sqrt{\lambda_i\lambda_j}
                                             \langle v_i | v_j \rangle
                                             | i \rangle\!\langle j | \right) | t \rangle \\
     &= \sum_{ij\in V} \sqrt{\lambda_i\lambda_j}
                       \, \overline{t_i}t_j \langle v_i | v_j \rangle \\
     &= \tr T\Lambda.
\end{split}\end{equation}
for an appropriate vector $| t \rangle \in \mathbb{C}^V$, because
the Hermitian matrices in the second and third line (the latter a Gram matrix)
have the same spectrum. The matrices $T$ and $\Lambda$ in the last line are
defined as follows:
\begin{align*}
  \Lambda_{ij} &= \sqrt{\lambda_j\lambda_i}, \\
  T_{ij}       &= \overline{t_i}t_j \langle v_i | v_j \rangle.
\end{align*}
When varying over quantum models of $G$ and states $\psi$, the
matrix $T$ varies over all semidefinite $T \geq 0$ such that
$\tr T = 1$ and $T_{ij} = 0$ whenever $i\sim j$ are connected
by and edge in $G$. \emph{I.e.},
\begin{equation}\begin{split}
  \vec{\lambda}(\mathcal{E}_{\mathrm{QM}}(\Gamma))
                &= \max \tr\Lambda T \\
                &\phantom{==}
                 \text{s.t. } T\geq 0,\ \tr T = 1,\ i\!\sim\! j \Rightarrow T_{ij}=0,
  \label{eq:primal-SDP}
\end{split}\end{equation}
which is indeed a semidefinite program.
\qed

\medskip
Closing this semidefinite discussion, the above primal SDP above has a dual,
as follows:
\begin{equation}\begin{split}
  \vec\lambda(\mathcal{E}_{\mathrm{QM}}(G))
     &= \min\, s \text{ s.t. } s\1 \geq S,\ S=S^\dagger, \\
     &\phantom{=====}
           (i\!\not\sim\! j \text{ or } i\!=\!j) \Rightarrow S_{ij}=\Lambda_{ij}.
  \label{eq:dual-SDP}
\end{split}\end{equation}
The value $\vec{\lambda}(\mathcal{E}_{\mathrm{QM}}(\Gamma))$
is known as a \emph{weighted Lov\'{a}sz number}
(or \emph{$\vartheta$-function})~\cite{Knuth94}.

\medskip
The previous discussion implies that not only function
optimization, but also membership in
$\mathcal{E}_\mathrm{QM}(\Gamma)$ is an efficient convex
problem: there is a polynomial-time algorithm that, given a
vector $\vec{p}$, tests whether it is in
$\mathcal{E}_\mathrm{QM}(\Gamma)$ or not. This follows from
general considerations of convex optimisation \cite{GLS-book,
BertsimasVempala, Liu}.

Does there exist such a nice and efficient description also for
the classical set $\mathcal{E}_\mathrm{C}(\Gamma)$? The fact
that the maximum of $\beta$ over it is the independence number
$\alpha(G)$, which is well-known to be NP complete, means that
the answer is ``no.'' In fact, $\mathcal{E}_\mathrm{C}(\Gamma)$
encodes the independence numbers $\alpha(G|_S)$ of all induced
subgraphs of $G$ on subsets $S \subset V$, and the best
description that we have is as the following $0$-$1$-polytope:
\begin{equation}
  \mathcal{E}_\mathrm{C}(\Gamma)
      = \operatorname{conv}\bigl\{ \vec{\sigma} : \sigma_i \in \{0,1\},\
                                                  i \!\sim\! j \Rightarrow \sigma_i\sigma_j = 0 \bigr\}.
\end{equation}

\medskip
Turning to generalized probabilistic models, $\beta_{\mathrm{GPT}}(\Gamma)$
seems at first much harder to characterize,
and we need to look at the full hypergraph structure.
Indeed, it is this value that we should with good
reason consider as the ``algebraic bound'' for $\beta$. After all,
it is the largest value we can assign to it under the most
general interpretation of the events $i \in V$ in a probabilistic model
that obeys the Gleason property.

The difficulty in evaluating $\beta_{\mathrm{GPT}}(\Gamma)$ lies in
capturing the constraint that the $P_i$ have to be sums of extremal,
normalized effects in the generalized probabilistic theory. If we
relax this condition simply to $P_i$ having to be an effect, we
arrive at what we would like to call a \emph{fuzzy model},
which formalizes the notion that all $\{ P_i : i\in C \}$ are
compatible, but not necessarily exclusive events: so we are left
with Gleason's constraints $0 \leq \langle P_i \rangle \leq 1$
and for all $C\in\Gamma$, $\sum_{i\in C} \langle P_i \rangle \leq 1$.
Denote the (convex) set of all expectations
$\bigl(\langle P_i \rangle\bigr)_{i\in V}$ when varying over
models and their states by $\mathcal{E}_{\mathrm{F}}(\Gamma)$.
%Vectors $\vec{p} \in \mathcal{E}_{\mathrm{F}}(\Gamma)$ are also known as
%\emph{fractional packings}. Then,
\begin{equation}
  \label{eq:beta-GPT-F}
  \beta_{\mathrm{GPT}}(\Gamma) = \beta_{\mathrm{F}}(\Gamma) = \alpha^*(\Gamma),
\end{equation}
where $\alpha^*(\Gamma)$ is the so-called \emph{fractional packing number}
of the hypergraph $\Gamma$, defined by the following intuitive
linear program:
\begin{equation}\begin{split}
  \alpha^*(\Gamma) &= \max \sum_{i\in V} w_i \\
                   &\phantom{==}
                      \text{s.t. } \forall i\ 0\leq w_i\leq 1 \text{ and }
                                   \forall C\in\Gamma\ \sum_{i\in C} w_i \leq 1.
\end{split}\end{equation}
The vectors $\vec{w}$ are known as \emph{fractional packings of
$\Gamma$}. To prove Eq.~(\ref{eq:beta-GPT-F}), observe on the
one hand that, for given fuzzy model $\{ P_i \}$ and a state
$\rho$, the weights $w_i = \langle P_i \rangle$ form a
fractional packing. Furthermore, a fractional packing $\{ w_i
\}$ \emph{is} a fuzzy noncontextual model for the unique
generalized probabilistic theory in $\mathbb{R}$, with the
usual ordering and unit $1$; the state is the identity. (In
other words, $\mathcal{E}_{\mathrm{F}}(\Gamma)$ is precisely
the polytope of fractional packings of $\Gamma$.)

Conversely, given a fractional covering $\vec{w}$, we now show
that there is an appropriate generalized probabilistic model
with effects $P_i$ and a state, such that $w_i = \langle P_i
\rangle$. Indeed, as the set of normalized states we choose
$\mathcal{S} = 1 \oplus \mathcal{E}_\mathrm{F}(\Gamma)$,
spanning a cone $\mathbb{R}_{\geq 0}\mathcal{S} \subset
\mathbb{R} \oplus \mathbb{R}^V$. The dual cone (with respect to
the usual Euclidean inner product) is the set of positive
observables: $\mathcal{S}' =: \mathcal{P} \subset \mathbb{R}
\oplus \mathbb{R}^V$ with unit element $u = 1 \oplus 0^V \in
\mathcal{P}$, which is $1$ precisely on the affine hyperplane
spanned by $\mathcal{S}$. Now, for each $i\in V$, let $P_i = 0
\oplus \delta_i \in \mathcal{P}$ be the $i$-th standard basis
vector. Clearly, for given fractional covering (\emph{i.e.},
state) $\vec{w}$ and all $i \in V$, $\langle P_i \rangle =
w_i$. Hence, all that remains to show is that these $P_i$ and
all $Q_C = u-\sum_{i\in C} P_i$ are extremal and normalized
(assuming that $C\in\Gamma$ is a maximal element). Concerning
normalization, observe that the fractional packings $\delta_i$
and $0$ (the all-zero assignment) yield proper states.
Regarding extremality, observe that on $S$, $P_i$ and all $Q_C$
are non-negative; furthermore, the equations $\langle P_i
\rangle = 0$ and $\langle Q_C \rangle = 0$ each define
hyperplanes intersecting $\mathbb{R}_{\geq 0}\mathcal{S}$ in a
convex set of dimension $|V|$, \emph{i.e.}~these equations
define facets of the cone $\mathbb{R}_{\geq 0}\mathcal{S}$,
meaning that all $\mathbb{R}_{\geq 0}P_i$ and $\mathbb{R}_{\geq
0}Q_C$ are indeed extremal rays.

Note that by the above argument we proved in fact that
$\mathcal{E}_{\mathrm{GPT}}(\Gamma) = \mathcal{E}_{\mathrm{F}}(\Gamma)$,
the set of fractional packings. This means that \emph{any}
linear function of expectation values can be optimized over
$\mathcal{E}_{\mathrm{GPT}}(\Gamma)$ as a linear program;
likewise, checking whether $\vec{p}$ is in $\mathcal{E}_{\mathrm{GPT}}(\Gamma)$
is a linear programming feasibility problem.
\qed

\medskip
For an example, for the $n$-cycles above, $\alpha^*(C_n) = n/2$,
regardless of the parity of $n$, which is strictly larger than
$\vartheta(C_n)$ for all odd $n \geq 5$. Again, we know of arbitrarily
large separations: there are hypergraphs $\Gamma$ such that the
adjacency graph $G$ is the complete graph $K_n$, hence
$\alpha(G) = \vartheta(G) = 1$,
yet $\alpha^*(\Gamma) \gg 1$~\cite{CLMW-0}.

\medskip\noindent
\textbf{Remark:} Our $\mathcal{E}_\mathrm{QM}(\Gamma)$ equals Knuth's set
${\tt TH}(\overline{G})$~\cite{Knuth94} for the adjacency graph $G$ of $\Gamma$;
likewise our $\mathcal{E}_\mathrm{C}(\Gamma)$ equals his ${\tt STAB}(G)$
and if $\Gamma$ is the hypergraph of all cliques in $G$, also
$\mathcal{E}_\mathrm{GPT}(\Gamma) = {\tt QSTAB}(G)$. Knuth introduced
these sets in his treatment of the (weighted) Lov\'{a}sz $\vartheta$-function,
independence numbers and fractional packing numbers, in an attempt to
explain the so-called ``sandwich theorem'' structurally.

%Of course we can also look at other parameters, as long as they
%make sense as expectation values of well-defined variables. This
%requires that the parameter in question can be written as a
%function of expectation values of functions of the $P_i$ for
%$i\in C$:
%\begin{equation}
%  \widetilde\beta = F\bigl(\langle g_C(P_i:i\in C) \rangle : C\in\Gamma \bigr).
%\end{equation}
%But for the moment we stick with the above choice.

\medskip%\noindent
{\it Bell inequalities.}---Where does nonlocality come into
this? After all, Bell inequalities exploit locality in the form
that one party's measurement is compatible with another
party's, and that the former's outcomes are independent of the
latter's choices (\emph{i.e.}, insensitive to different
contexts). We can model this also in our setting, by going to
the atomic events, which are labelled by a list of settings and
outcomes for each party. For instance, for bipartite scenarios,
let Alice and Bob's settings be $x\in\mathcal{X}$ and
$y\in\mathcal{Y}$, respectively, and their respective outcomes
be $a\in\mathcal{A}$ and $b\in\mathcal{B}$. Then, we construct
a graph with vertex set $V =
\mathcal{A}\times\mathcal{B}\times\mathcal{X}\times\mathcal{Y}$
and edges $abxy \sim a'b'x'y'$ if and only if ($x=x'$ and
$a\neq a'$) or ($y=y'$ and $b\neq b'$), encoding precisely that
two events in $V$ are connected in the graph if and only if
they are compatible and mutually exclusive (as events in the
Bell experiment as a whole). Let $\Gamma$ be the hypergraph of
all cliques in $G$.

We can now discuss classical noncontextual, quantum and
generalized models for this graph, and hence also noncontextual
inequalities, restricting as above to linear functions
$\vec\lambda \dot \vec{p}$ of the vector of the probabilities
$p_{ab|xy} = \langle P_{abxy} \rangle$, with with non-negative
coefficient vector $\vec\lambda$. Note that any Bell inequality
can always be rewritten in such a form, by removing negative
coefficients using the identity $-p_{ab|xy} = -1+\sum_{a'b'\neq
ab} p_{a'b'|xy}$ for all $x$, $y$, $a$, and $b$. These
equations are not automatically realized in the sets
$\mathcal{E}_{\mathrm{X}}(\Gamma)$,
$\mathrm{X}=\mathrm{C,QM,GPT}$ -- as indeed in the underlying
(classical, quantum or generalized) model it needs not hold
that $\sum_{ab} P_{abxy}$ is the unit element, for any $x,y$.
Hence, define for any class of models
$\mathrm{X}=\mathrm{C,QM,GPT}$,
\begin{equation}
  \mathcal{E}^{1}_{\mathrm{X}}(\Gamma) := \mathcal{E}_{\mathrm{X}}(\Gamma)
                                           \cap
                                          \left\{ \vec{p} : \forall xy\ \sum_{ab} p_{ab|xy} = 1 \right\},
\end{equation}
the set of probability assignments consistent with the contextuality
structure $\Gamma$, and in addition satisfying normalization.

In the appendix we prove (which is not too difficult) that
$\mathcal{E}^{1}_{\mathrm{C}}(\Gamma)$ is precisely the set of correlations
explained by local hidden variable theories, and that
$\mathcal{E}^{1}_{\mathrm{GPT}}(\Gamma)$ are exactly the no-signalling
correlations.
Furthermore, to calculate the local hidden variable value $\Omega_c$
of a given Bell inequality with non-negative coefficient vector $\vec\lambda$,
it holds that
\begin{equation}
  \Omega_c = \vec\lambda(\mathcal{E}^{1}_{\mathrm{C}}(\Gamma))
           = \vec\lambda(\mathcal{E}_{\mathrm{C}}(\Gamma)).
\end{equation}
In this sense, any Bell inequality is at the same time a noncontextual
inequality for the underlying graph $G$.

With classical and no-signalling correlations taken care of,
we turn our attention to the quantum case. Once again, we refer
the reader to the appendix for a proof that the following subset
of $\mathcal{E}^{1}_{\mathrm{QM}}(\Gamma)$ is precisely the set of
correlations obtainable by local quantum measurements on a bipartite
state (where ``local'' means that all operators of one party
commute with all operators of another party):
\begin{equation}
  \mathcal{E}^{\1}_{\mathrm{QM}}(\Gamma)
      = \left\{ \bigl(\langle P_{abxy} \rangle\bigr)_{abxy}
                   : \forall xy\ \sum_{ab} P_{abxy} = \1 \right\}.
\end{equation}
\emph{I.e.}, we add the completeness relation for the measurements
in the model. This of course also means that for a given Bell
inequality with coefficients $\vec\lambda$, the maximum quantum
value is
\begin{equation}
  \Omega_q = \vec\lambda(\mathcal{E}^{\1}_{\mathrm{QM}}(\Gamma)).
\end{equation}

For the time being we do not know whether the set of quantum
correlations, \emph{i.e.}~$\mathcal{E}^{\1}_{\mathrm{QM}}(\Gamma)$,
is efficient to characterize. It follows, however, from the
above considerations and the general theory of convex
optimization~\cite{GLS-book,BertsimasVempala,Liu}
that the -- potentially larger -- set
$\mathcal{E}^{1}_{\mathrm{QM}}(\Gamma)$ can be decided efficiently.
In fact, we shall see directly that the maximum values
$\vec\lambda(\mathcal{E}^{1}_{\mathrm{QM}}(\Gamma))$ are
computed to arbitrary precision by semidefinite programming,
thus providing efficient upper bounds to $\Omega_q$.

Namely, for $M \gg 1$, consider the linear function
\begin{equation}
  \vec\lambda \cdot \vec{p} + M \sum_{xy} \biggl( \! -1 \!+\! \sum_{ab} p_{ab|xy} \! \biggr)
        = \bigl( \vec\lambda + M\vec{1} \bigr) \cdot \vec{p} - M|\mathcal{X}\times\mathcal{Y}|,
\end{equation}
which encodes $\vec\lambda$ plus a large negative penalty for
any $xy$ such that $\sum_{ab} p_{ab|xy} < 1$, and maximize it
over the full set of quantum models, $\mathcal{E}_{\mathrm{QM}}(\Gamma)$.
[Note that ``$\leq 1$'' is guaranteed by the Gleason property, which
is valid in this set.]
Clearly, all the values
$(\vec\lambda+M\vec{1})(\mathcal{E}_{\mathrm{QM}}(\Gamma))$ are instances
of the semidefinite programs discussed earlier, and as $M\rightarrow\infty$,
\begin{equation}
  (\vec\lambda+M\vec{1})(\mathcal{E}_{\mathrm{QM}}(\Gamma)) - M|\mathcal{X}\times\mathcal{Y}|
     \longrightarrow \vec\lambda(\mathcal{E}^{1}_{\mathrm{QM}}(\Gamma)).
\end{equation}

Implementing this for example for the CHSH inequality~\cite{CHSH69}, we recover the
Tsirelson bound $2\sqrt{2}$~\cite{Tsirelson80} -- see the appendix for details.
On the other hand, for the $\mathrm{I}_{3322}$ inequality~\cite{BG07} the method
yields the upper bound $0.251\, 47$ on the quantum value;
the currently best upper bound is slightly smaller:
$0.250\, 875\, 56$~\cite{PV10}, from which we conclude that in general,
$\mathcal{E}^{\1}_{\mathrm{QM}}(\Gamma)$ is strictly contained in
$\mathcal{E}^{1}_{\mathrm{QM}}(\Gamma)$ -- once more, see the appendix for details.
[As an aside, we note that in the latter case, maximizing over
$\mathcal{E}_{\mathrm{QM}}(\Gamma)$ gives the even much larger
bound $0.4114$ -- so, unlike classical models, in the quantum
the probability normalization is not for granted.]

\medskip%\noindent
{\it Conclusions.}---Notice that the previous exposition bears
striking similarity to the discussion of the no-signalling
property in the context of classical, quantum, or more general
correlations. Indeed, as it was observed by Popescu and
Rohrlich \cite{PR94}, and Tsirelson \cite{Tsirelson93}, the
no-signalling principle is not enough to explain the scope of
quantum correlations; for instance, for the CHSH inequality,
the classical bound is $2$, the quantum bound is $2\sqrt{2}$,
while the algebraic bound $4$ is attainable under the most
general no-signalling correlations. Likewise here: operational
models obeying the Gleason constraint include classical and
quantum ones, but they definitely go beyond these two. One
might ask: why is nature not even more contextual than quantum
mechanics?

Unlike Bell inequalities, here we see that the maximum quantum
violation is always efficiently computable, as it is the
solution to a semidefinite program, and these are solvable in
polynomial time. Thanks to the general machinery of convex
optimisation problems \cite{GLS-book,BertsimasVempala,Liu},
this also means that membership of a probability assignment
$\vec{p}$ in $\mathcal{E}_\mathrm{QM}(\Gamma)$ can be tested
efficiently, despite the fact that the set is not itself
defined directly by semidefinite constraints. Generalized
models are captured instead entirely by linear inequalities and
linear programming -- in particular, also here all maximum
violations of noncontextual inequalities can be computed
efficiently, as linear programs.
At the other end of the spectrum, the
noncontextual set $\mathcal{E}_\mathrm{C}(\Gamma)$ is the
convex hull of many, but easy to describe points, but its
characterisation in terms of inequalities is computationally
hard, and so are maximum values such as
$\beta_\mathrm{C}(\Gamma)$, which can be as hard as NP
complete.

The sets of probability assignments compatible with
noncontextual, quantum and generalized operational models are
different from each other even in the simplest nontrivial case,
that of the pentagon, as witnessed by the values $2$,
$\sqrt{5}$, and $5/2$ for $\beta(5)$, respectively. Especially
the gap between $\sqrt{5}$ for quantum and $5/2$ for
generalized models is noteworthy, because the latter value is
attained by putting weight $1/2$ to each vertex in a Gleason
assignment of probabilities to each of the five vertices of
$C_5$. It had been noted by other authors before, that the
Gleason constraint on finite sets of vectors allows assignments
incompatible with quantum theory \cite{Wright78}. We believe
that here we clarified this observation further, since we
showed that each such assignment originates in fact from a
sound operational model based on generalized probabilistic
theories. Each vertex is assigned an event such that, with
respect to the given state, any adjacent pair is ``complete''
in the sense that the probabilities add up to $1$. It is easy
to see that quantum mechanics cannot yield this, as it would
require successive subspace projectors to be orthogonal
complements of each other.

We close by highlighting some open questions:
Looking back, it is the insistence on exclusiveness of events,
and the dropping of completeness relations, that made the
KCBS inequalities and our generalizations possible; not insisting
on effects having to sum to unity (always prominent in the
``usual'' KS proofs) also seems responsible for
the fact that we obtain a semidefinite program for the maximum
quantum value. On the other hand, how to incorporate this as
an additional constraint in the SDP?

As this seems to mark exactly the difference between nonlocal
quantum values and quantum violations of generalized KCBS
inequalities, the question arises: how good is the latter as a
bound on the former? And how does it relate to upper bounds
obtained from the Navascu\'{e}s-Pironio-Ac\'{\i}n hierarchy
\cite{NPA08}?

%%%%%%%%%%%%%%%%%%%%%%%%%%%%%%%%%%%%%%%%%%%%%%%%%%%%%%%%%%%%%%%%%%%

\bigskip\noindent
{\it Acknowledgments.}---We thank P.\ Badzi{\c a}g, J.\
Barrett, I.\ Bengtsson, T.\ Cubitt, A.\ Harrow, A.\ Klyachko,
D.\ Leung, J.-{\AA}.\ Larsson, W.\ Matthews, J.\ Oppenheim, and
K.\ Svozil for conversations.

%We thank Piotr Badzi{\c a}g, Jonathan Barrett, Ingemar Bengtsson,
%Toby Cubitt, Aram Harrow, Alexander Klyachko, Debbie Leung, Jan-{\AA}ke Larsson,
%William Matthews, Jonathan Oppenheim, and Karl Svozil for
%conversations.

The present research was supported by the European Commission,
the U.K.~EPSRC, the Royal Society, the British Academy, the
Royal Academy of Engineering, the Spanish MCI Project No.\
FIS2008-05596, and by the National Research Foundation as well
as the Ministry of Education of Singapore.

%%%%%%%%%%%%%%%%%%%%%%%%%%%%%%%%%%%%%%%%%%%%%%%%%%%%%%%%%%%%%%%%%%%

%\bigskip

\appendix

%%%%%%%%%%%%%%%%%%%%%%%%%%%%%%%%%%%%%%%%%%%%%%%%%%%%%%%%%%%%%%%%%%%

\section{Non-locality: proofs}
Here we prove the claims in the Bell inequality section.

\medskip\noindent
\emph{(i) Proof that $\mathcal{E}^{1}_{\mathrm{C}}(\Gamma) =$
local realistic correlations.} If the $A$'s and $B$'s form a
(deterministic) classical local hidden variable model, then the
products $P_{xy}^{ab} = A_x^a B_y^b$ are a classical
noncontextual model for the graph $G$. Since for each $x$ and
$y$ there is exactly one $a$, $b$, respectively, such that
$A_x^a=B_y^b=1$, the normalization condition is fulfilled, too.

Vice versa, given any deterministic noncontextual model
$P_{xy}^{ab}$ for $G$ we show how to construct local hidden
variables $A_x^a$ and $B_y^b$ (taking values $0$ and $1$) such
that $P_{xy}^{ab} \leq A_x^a B_y^b$; using the probability
normalization, this must be an equality.
Namely, assume $P_{abxy} = 1$ for any quadruple $abxy$. Then, thanks to the
graph $G$, for any $a'\neq a$ and any $y$ and $b$, $P_{a'b'xy'} = 0$. In other
words, for every $x$, there is at most one $a$ such that $P_{ab'xy'} = 1$
for any $b'y'$. Choose this $a$ (or else an arbitrary one) to let
$A_x^a=1$ and all other $A_x^{a'}=0$. Likewise for $B_y^b$, and we clearly
obtain the claim.
\qed

\medskip\noindent
\emph{(ii) Proof that $\mathcal{E}^{1}_{\mathrm{GPT}}(\Gamma) =$ no-signalling correlations.}
Let $\vec{p} \in \mathcal{E}_{\mathrm{GPT}}(\Gamma)$ such that for all $xy$,
$\sum_{ab} p_{ab|xy} = 1$. We have to show the no-signalling relations,
\begin{align*}
  \forall ax\forall yy' \quad \sum_b p_{ab|xy} &= \sum_b p_{ab|xy'}, \\
  \forall by\forall xx' \quad \sum_a p_{ab|xy} &= \sum_a p_{ab|x'y}.
\end{align*}
To prove this, note for fixed $x$, $y$ and $y'$, that the vertices
\[
  \{ abxy : b\in\mathcal{B} \} \cup \{ a'bxy' : a'\in\mathcal{A}\setminus a,\, b\in\mathcal{B} \}
\]
form a clique in $G$, hence
\[
  \sum_{b} p_{ab|xy} + \sum_{a'\neq a, b} p_{a'bxy'} \leq 1,
\]
which implies $\sum_{b} p_{ab|xy} \leq \sum_{b} p_{ab|xy'}$ for arbitrary $y$
and $y'$. By symmetry, equality must hold.
\qed

\medskip\noindent
\emph{(iii) Proof that
$\vec\lambda(\mathcal{E}^{1}_{\mathrm{C}}(\Gamma)) =
\vec\lambda(\mathcal{E}_{\mathrm{C}}(\Gamma))$.} Recall from
(i) that we can find, for any deterministic noncontextual model
$P_{xy}^{ab}$, local hidden variables $A_x^a$ and $B_y^b$
(taking values $0$ and $1$) such that $P_{xy}^{ab} \leq A_x^a
B_y^b$. The right hand side is in evidently in
$\mathcal{E}^{1}_{\mathrm{C}}(\Gamma))$.
Hence, for the purpose of maximizing
a objective function with non-negative coefficients $\vec{\lambda}$,
we may restrict to $\mathcal{E}^{1}_{\mathrm{C}}(\Gamma))$.
\qed

\medskip\noindent
\emph{(iv) Proof that $\mathcal{E}^{\1}_{\mathrm{QM}}(\Gamma)
=$ quantum correlations.} We face a problem like in (i): given
operators $P_{abxy}$ forming a quantum model of $G$, we have to
define projector valued measurements
$(A_x^a)_{a\in\mathcal{A}}$ and $(B_y^b)_{b\in\mathcal{B}}$
such that $[A_x^a,B_y^b]=0$ and $P_{abxy} = A_x^a B_y^b$.

There are obvious candidates for these ``local'' measurements given as marginals
of $P_{abxy}$:
\begin{align*}
  A_x^a &= \sum_{b'} P_{ab'xy} \quad (\text{for any } y), \\
  B_y^b &= \sum_{a'} P_{a'bxy} \quad (\text{for any } x),
\end{align*}
which raises the immediate issue that, \emph{a priori}. the
right hand sides may not be independent of $y$ and $x$,
respectively. Denote the right hand sides above by $A_{xy}^a$
and $B_{xy}^b$. We show that the assumption of completeness,
$\sum_{ab} P_{abxy} =\1$, implies that $A_{xy}^a$ is
independent of $y$, $B_{xy}^b$ independent of $x$.
Indeed, observe that for any $a'\neq a$ and any $y$, $y'$, $b$,
and $b'$, $P_{abxy} \perp P_{a'b'xy'}$, which by summation
implies that
\[
  A_{xy}^a \perp \sum_{a'\neq a} A_{xy'}^{a'} = \1 - A_{xy'}^a,
\]
\emph{i.e.}, $A_{xy}^a \leq A_{xy'}^a$ for all $y$ and $y'$. By symmetry
we hence must have $A_{xy}^a = A_{xy'}^a$ and likewise
$B_{xy}^b = B_{xy'}^b$.

Now, observe finally
\[
  A_x^a B_y^b = \sum_{a' b'} P_{ab'xy}P_{a'bxy} = P_{abxy} = B_y^b A_x^a,
\]
and we are done.
\qed

\medskip\noindent
\emph{(v) Example CHSH.}
Here, $\mathcal{A}=\mathcal{B}=\mathcal{X}=\mathcal{Y}=\{0,1\}$
and $\vec{\lambda}$ encodes the winning condition for the CHSH (or PR) game:
\begin{equation}
  \lambda_{abxy} = \begin{cases}
                     1 & :\ a\oplus b = xy, \\
                     0 & :\ \text{otherwise.}
                   \end{cases}
\end{equation}
The CHSH inequality expresses the fact that $\Omega_{\mathrm{c}} = 3$
while $\Omega_{\mathrm{q}} = 2+\sqrt{2}$.

Constructing the graph and the matrices $\Lambda$ and $T$ by
hand is easy: $G$ has $16$ vertices, so the matrices are also
$16 \times 16$. Since $\Lambda$ is rather sparse, this allows
us immediately to reduce it to a graph $G'$ on $8$ vertices
with new $\Lambda$-matrix equal to $J$, the all-$1$-matrix. The
graph is the $(1,4)$-circulant graph on $8$ vertices; one can
obtain it by joining antipodal vertices in the $8$-cycle $C_8$.
So, we find that $\lambda(\mathcal{E}_{\mathrm{QM}}(G)) =
\vartheta(G')$, and the latter is easily evaluated to
$2+\sqrt{2}$, using the dual characterisation of Lov\'{a}sz
(\emph{i.e.}~our dual SDP).

\medskip\noindent
\emph{(vi) Example $\mathbf{\text{I}_{3322}}$.}
This is a Bell inequality for $3$ settings for each Alice and Bob,
each measurement having binary output. In the form found in~\cite{BG07}
it reads
\begin{equation}
\begin{split}
  -2 &\langle A_0^0 \rangle      - \langle A_1^0 \rangle        - \langle B_0^0 \rangle         \\
     &+ \langle A_0^0 B_0^0 \rangle + \langle A_0^0 B_1^0 \rangle + \langle A_0^0 B_2^0 \rangle \\
     &+ \langle A_1^0 B_0^0 \rangle + \langle A_1^0 B_1^0 \rangle - \langle A_1^0 B_2^0 \rangle \\
     &+ \langle A_2^0 B_0^0 \rangle - \langle A_2^0 B_1^0 \rangle
       \leq 0,
\end{split}
\end{equation}
and the value $0$ is the maximum attainable under local hidden
variables. One form of the objective function with non-negative
coefficients, using the above substitution trick, is
$\vec{\lambda}\cdot\vec{p}$, with the vector $\vec{\lambda} \in
\mathbb{R}^{36}$ being given by the following table:
\smallskip
\begin{center}
\begin{tabular}{l|cc|cc|cc}
  $xa \setminus yb$ & 00 & 01 & 10 & 11 & 20 & 21 \\
  \hline
                 00 &  1 &  0 &  1 &  0 &  1 &  0 \\
                 01 &  0 &  0 &  1 &  1 &  1 &  1 \\
  \hline
                 10 &  1 &  1 &  1 &  0 &  0 &  1 \\
                 11 &  0 &  1 &  1 &  1 &  1 &  1 \\
  \hline
                 20 &  1 &  0 &  0 &  1 &  0 &  0 \\
                 21 &  0 &  0 &  1 &  1 &  0 &  0
\end{tabular}
\end{center}
\smallskip
The classical bound is $\Omega_{\mathrm{c}} = 6$,
while the best known quantum violation attains a value $6.250\,
875\, 384 \leq \Omega_{\mathrm{q}}$; on the other hand, it is
known that $\Omega_{\mathrm{q}} \leq 6.250\, 875\, 56$, by
going as far up in the Navascu\'{e}s-Pironio-Ac\'{\i}n
hierarchy \cite{NPA08} as was computationally feasible (almost
the fourth level); the conjecture is that this is essentially
the optimal value, although there is still disagreement from
the 7th digit on. It is also conjectured that to attain the
quantum limit, an infinitely large entangled state is required
-- in~\cite{PV10} a candidate sequence of larger and larger
states and measurements is presented which give better and
better values suggested to converge to the optimum.

The game context graph $G$ on $36$ vertices in not constructed explicitly
here, though it is easy. Looking at the primal SDP, and noticing
that only $20$ out of $36$ components of $\vec{\lambda}$ are
populated, and then only by $1$'s, one sees -- cf.~the CHSH case -- that,
by constructing the induced subgraph $G'$ of the context graph on
the $20$ vertices $abxy$ with $\lambda_{abxy} = 1$, we obtain
$\vec\lambda(\mathcal{E}_{\mathrm{QM}}(G)) = \vartheta(G') \approx 6.4114$.

This is an instance of the
probabilities simply not adding up to $1$, in other words:
$\vec\lambda(\mathcal{E}^{1}_{\mathrm{QM}}(G))$ is strictly
smaller. Indeed, a calculation with on SeDuMi
resulted in
$\vec\lambda(\mathcal{E}^{1}_{\mathrm{QM}}(G)) \approx 6.251\, 47$.

%%%%%%%%%%%%%%%%%%%%%%%%%%%%%%%%%%%%%%%%%%%%%%%%%%%%%%%%%%%%%%%%%%%

\end{document}